\documentclass[prd,twocolumn,amsmath,amssymb,superscriptaddress,floatfix]{revtex4}
\usepackage{bm}
\usepackage{amsmath}
\usepackage{epsf}
\usepackage{color}
\usepackage{natbib}
\usepackage{graphicx}
\usepackage{multirow}
\usepackage{hyperref}
\usepackage{ifthen}

\newcommand{\rest}{{\rm (rest)}}

\def\simlt{\lesssim}

\newcommand{\ck}{c_K}
\newcommand{\tppi}{\ck p_T \pi_T}

\begin{document}
\title{Crossing the Phantom Divide with Parameterized Post-Friedmann Dark Energy}

\author{Wenjuan Fang}
\affiliation{Department of Physics, Columbia University, New York,
NY 10027}
\author{Wayne Hu}
\affiliation{Kavli Institute for Cosmological Physics, Astronomy \&
Astrophysics Department, Enrico Fermi Institute, University of Chicago,
Chicago, IL 60637}
\author{Antony Lewis}
\affiliation{Institute of Astronomy, Madingley Road, Cambridge, CB3 0HA, UK.}

\begin{abstract}
Dark energy models with a single scalar field cannot cross the
equation of state divide set by a cosmological constant. More
general models that allow crossing require additional degrees of
freedom  to ensure gravitational stability. We show that a
parameterized post-Friedmann description of cosmic acceleration
provides a simple but accurate description of multiple scalar field
crossing models.  Moreover the prescription provides a well
controlled approximation for a wide range of ``smooth" dark energy
models.  It conserves energy and momentum and is exact in the metric
evolution on scales well above and below the transition scale to
relative smoothness.  Standard linear perturbation tools have been
altered to include this description and made publicly available for
studies of the dark energy involving cosmological structure out to
the horizon scale.

\end{abstract}
\date{\today}
\maketitle

\section{Introduction} \label{sec:intro}

Observational constraints on the acceleration of the expansion have
continued to close in on a dark energy equation of state of a
cosmological constant $w_e=-1$ that delineates the phantom divide.
Testing the small deviations from that value in the future requires
a description of the dark energy that allows the equation of state
to evolve across the phantom divide possibly multiple times.

It is well known that single scalar fields are gravitationally
unstable to such a crossing of the phantom divide
\cite{Vik04,Hu04c,CalDor05}. Dark energy that is minimally coupled
to the matter requires additional degrees of freedom to cross the
divide stably.  While specific models with multiple fields can be
constructed \cite{Hu04c,Feng:2004ad} they are cumbersome or
impossible to implement in a general analysis of the dark energy.

The usual approach in the literature for finessing such cases is to
artificially turn off the dark energy perturbations explicitly or
implicitly by limiting the range of observables.  Doing so violates
energy-momentum conservation whenever $w_e \ne -1$ and leads to
inconsistencies between the Einstein equations for the evolution of
the metric due to the Bianchi identities which can persist even on
small scales. Though the impact of perturbations tend  to be small
near $w_e=-1$, cosmological constraints often require the
exploration of a large swath of parameter space around the maximum
likelihood.  Excising the instability around the transition provides
another, albeit rather ad hoc approach \cite{Hue05}.

In this {\it Brief Report}, we show that the so-called parameterized
post-Friedmann (PPF) approach to describing linear metric evolution
in a Friedmann-Robertson-Walker (FRW) universe  provides a simple
solution to this dilemma.  The PPF framework was ostensibly
introduced for describing modified gravity theories under a metric
framework with strict local conservation of energy and momentum
\cite{HuSaw07b}.  As such it also applies to dark energy models
\cite{Hu08a} and in particular the class of models which have  a
well-defined Jeans scale under which the dark energy is smooth
compared to the dark matter. This framework also has the benefit of
being an exact description for the metric evolution well above and
well below this scale and hence provides a very well-controlled
approximation that is simple to implement in an Einstein-Boltzmann
linear perturbation code.

\section{Phantom Divide and Scalar Fields}

Minimally coupled scalar field dark energy models that evolve across
the phantom divide require new internal degrees of freedom to
maintain gravitational stability.  To see this fact, consider the
conservation equation for the momentum density $(\rho_e {\bf u}_e)_i
\equiv T^0_{\hphantom{0}i}$ (see, e.g.~\cite{Hu08a} for an explicit
derivation)
\begin{equation}
 u_e' = (3w_e-1) u_e + k_H {\delta p_e \over\rho_e}  + (1+w_e) k_H A \,,
\end{equation}
where $' \equiv d/d\ln a$, $w_e=p_e/\rho_e$, $k_H = k/aH$, and $A$
is the gravitational potential in an arbitrary gauge.

The relationship between the pressure and density fluctuation
defines a sound speed. For a single scalar field with kinetic and
potential degrees of freedom, this relationship is most simply
described in a coordinate system that comoves with the dark energy
such that the momentum density and transverse spatial metric
fluctuations vanish\cite{Hu98,ArmMukSte00}.  From an arbitrary
gauge, this quantity is obtained by a gauge transformation that
changes the time slicing
\begin{eqnarray}
\delta \rho^\rest &=& \delta  \rho_e + 3\rho_e {u_e \over k_H}\,,  \nonumber\\
\delta p^\rest &=& \delta p_e + 3{p_e' \over \rho_e'}\rho_e {u_e \over k_H}\,,
 \end{eqnarray}
which defines a sound speed
\begin{equation}
c_s^2 \equiv  {\delta p_e^\rest \over \delta \rho_e^{\rest} }\,,
\end{equation}
bringing the momentum conservation equation to
\begin{equation}
 u_e' = 3\left( w_e+c_s^2-{p_e' \over \rho_e'}-{1\over 3}\right)
 u_e + k_H c_s^2 \delta_e + (1+w_e) k_H A\,, \nonumber
\end{equation}
where $\delta_e = \delta \rho_e /\rho_e$.

For a single scalar field, the rest or zero momentum gauge
corresponds to time slicing where the field, and hence the potential
energy, is constant leaving the energy density and pressure to be
defined by fluctuations in the kinetic energy. For a canonical
kinetic term $c_s^2 = 1$ representing the familiar kinetic energy
dominated equation of state of such scalars.

The dark energy system is completed by the continuity equation
\begin{eqnarray}
&&\delta_e' + 3(c_s^2-w_e) \delta_e + 9\left(c_s^2 - {p_e'\over \rho_e'} \right) {u_e \over k_H}
= \nonumber\\
&&\quad -k_H u_e - (1+w_e)(k_H B + 3 H_L')\,,
\end{eqnarray}
where $B$ is the space-time piece  and $H_L$ the space-space curvature
piece of the metric fluctuations in an arbitrary gauge \cite{Bar80}.

Taking $c_s^2>0$ makes dark energy perturbations Jeans stable in the
regime $k_H c_s \gg 1$  so long as $p_e'/\rho_e'$ remains finite.
In the matter dominated epoch, matter density fluctuations continue
to grow and so the Poisson equation for $\Phi \equiv H_L^{\rm
(newt)}$ in the Newtonian gauge
\begin{equation}
c_K k^2 \Phi = 4\pi G a^2 \sum_i \rho_i \delta_i^{\rest}
\label{eqn:poisson}
\end{equation}
becomes dominated by the matter component, {\it i.e.}\ the dark
energy is relatively smooth compared with the matter
\begin{equation}
\rho_e \delta_e^\rest \ll \rho_T \delta_T^\rest\,,
\label{eqn:smooth}
\end{equation}
where $``T"$ denotes all other components excluding the dark energy.
Here $c_K = 1-3K/k^2$ where $K$ is the background curvature.

This condition for smoothness is not the same as setting all dark
energy perturbations to zero which causes inconsistencies between
the four scalar Einstein equations. In particular, in the
synchronous gauge, where the dark matter momentum also vanishes,
some care must be taken even at $k_H c_s \gg 1$ since the dark
energy momentum is no longer negligible in comparison \cite{Hu08a}.

When $w_e=-1$, $p_e'/\rho_e'$ will generally diverge leading to an
instability in the evolution of perturbations if $c_s^2$ is held
fixed \cite{Hu04c}.  The problem arises since the change in the time
slicing required to reach the rest or constant field gauge becomes
infinite  when the field has no kinetic energy.  Viewed as a fluid,
the problem is that the relative fluid velocity $v_e = u_e/(1+w_e)$
becomes undefined if the momentum remains finite.

If the dark energy is a composite of fields then $c_s^2$ need not
itself be fixed by fundamental properties of the scalars at the
crossing.  For example if the dark energy were composed of the sum
of minimally coupled fields each with sound speed $c_e^2$ then the
pressure fluctuation is described by
\begin{eqnarray}
\delta p_e &=& c_s^2 \rho_e \delta_e + 3\left( c_s^2 - {p_e' \over
\rho_e'} \right) {\rho_e u_e \over k_H}
\nonumber\\
&=&   c_e^2 \rho_e \delta_e + 3\left( c_e^2 {\rho_e u_e \over k_H} -
\sum_\alpha {p_{e\alpha}' \over \rho'_{e\alpha}} {\rho_{e\alpha}
u_{e\alpha} \over k_H}  \right)\,,
\end{eqnarray}
which implicitly defines $c_s^2$ as a function of the individual
momenta.  As long as no individual component crosses the phantom
divide $w_{e\alpha}\ne-1$, the pressure fluctuations are no longer
singular.

Simple two field models which cross $w_e=-1$ were constructed in
\cite{Feng:2004ad,Hu04c}. Unfortunately, this  construction is
cumbersome for obtaining a general function $w_e(\ln a)$ constrained
to match cosmological distances.

The spirit of this construction is more broadly applicable.  Models
that cross the phantom divide must have internal degrees of freedom
to ensure $u_e$ remains finite through the crossing.  Provided they
do, energy momentum conservation and the requirement that the dark
energy is smooth compared with the matter for $c_e k_H \gg 1$ impose
nearly unique constraints on their parameterization. We will use
these requirements to construct a PPF description of dark energy
crossing.

\section{PPF Description}

The PPF description of dark energy replaces the density and momentum
components with a single joint dynamical variable $\Gamma$ but
retains strict conservation of energy and momentum in its equation
of motion.

Given the conservation laws, PPF and more generally any minimally
coupled dark energy parameterization requires two closure conditions
to complete the system \cite{Hu98}. The first can be taken as a
condition on the anisotropic stress.  For scalar fields this
quantity vanishes for linear field perturbations.

In the discussion above, the second condition was taken to be the
relationship between pressure and density fluctuations.  We saw that
this choice leads to difficulties in parameterizing models that
cross the phantom divide due to the appearance of singularities in
the equation of motion for the momentum density.

The PPF description replaces this condition on the pressure
perturbations with a direct relationship between the momentum
density of the dark energy and that of matter on large scales and a
transition scale under which the dark energy explicitly becomes
relatively smooth. The latter implicitly describes the momentum
density on small scales. The strategy for choosing these
relationships is to match the evolution of the metric exactly for
scales much larger and much smaller than the transition scale.

Let us start with the $\Gamma$ variable.  The conditions that the
anisotropic stress of the dark energy vanishes and the Poisson
equation is normal on small scales reduces the defining equation to
(see \cite{Hu08a}  Eq.  30)
\begin{equation}
 \Gamma \equiv { 4\pi G a^2 \over c_{K}k^{2}} \rho_T \Delta_T -  \Phi \,,
 \label{eqn:gamma}
\end{equation}
where
\begin{equation}
\Delta_T \equiv  \delta_T^{\rest} = \delta_T + 3 u_T/k_H
\end{equation}
is the density fluctuation in the zero momentum (total matter or
comoving) gauge of the matter excluding the dark energy.  Comparing
this relationship with the Poisson equation (\ref{eqn:poisson})
yields
\begin{equation}
\Gamma = - {4\pi G a^2 \over k^2 c_K}\rho_e \delta_e^{\rest} \,.
\end{equation}
The condition that the dark energy becomes smooth relative to the matter in
their respective rest gauges then becomes a direct requirement on the
evolution of $\Gamma$.

 \begin{figure}[t]
\begin{center}
\includegraphics[width=3.4in]{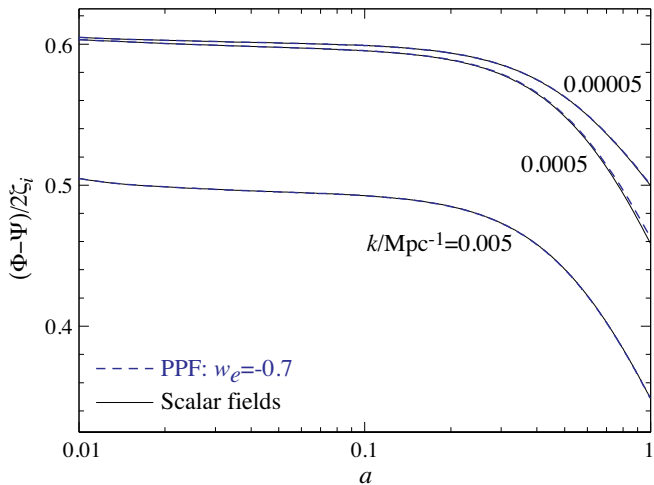}
\caption{PPF vs scalar field calculation of the evolution of the
potential responsible for gravitational redshifts and lensing
$(\Phi-\Psi)/2$ for a $w_e=-0.7$ model (flat, with $\Omega_m=0.31$
and $h=0.64$). Curves are normalized to the initial curvature
$\zeta_i$.} \label{fig:timeplot}
\end{center}
\end{figure}

\begin{figure}[t]
\begin{center}
\includegraphics[width=3.4in]{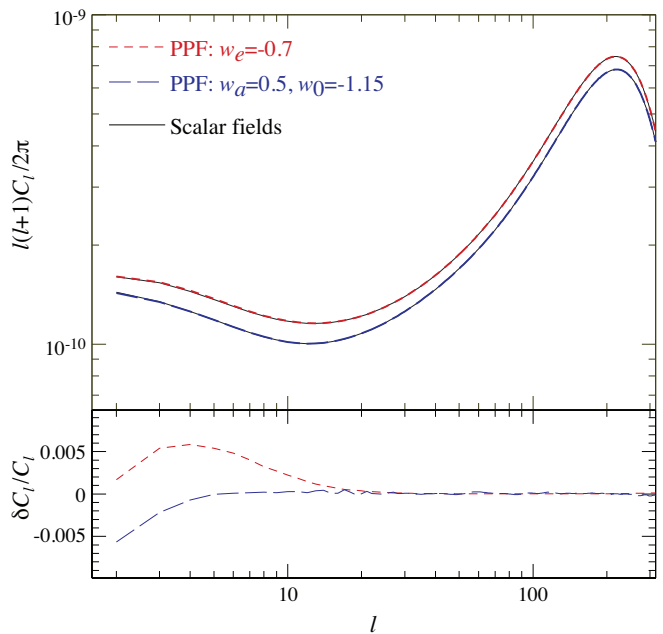}
\caption{PPF vs scalar field calculation of the CMB anisotropy power
spectrum for the $w_e=-0.7$ model of Fig.~\ref{fig:timeplot} and a
two-field crossing model that approximates $w_0=-1.15$ and $w_a=0.5$
(flat, with $\Omega_m=0.26$ and $h=0.74$). } \label{fig:clplot}
\end{center}
\end{figure}

Now let us examine the second closure relation.  On large scales,
energy and momentum conservation determine that  the curvature
$\zeta \equiv H_L^{(T)}$ in the total matter gauge is conserved up
to order $k_H^2$ in a flat universe with adiabatic fluctuations
\cite{Bar80}. The corresponding evolution equation for the Newtonian
potentials $\Phi$ and $\Psi$ is closed by the anisotropic stress
assumption \cite{HuEis99,Ber06}.

The Einstein equation governing  $\zeta$ reads
\begin{eqnarray}
\zeta' &=& \xi - {K \over (aH)^2}{V_T \over k_H}
- {4\pi G \over H^2} \rho_e {U_e  \over k_H}\,,
\label{eqn:zetaprimegeneral}
\end{eqnarray}
where $V_T=B^{(T)}$, $\xi=A^{(T)}$ and $U_e = u_e^{(T)}$ in this gauge and
\begin{align}
\xi =  -{\Delta p_T - {2\over 3}\tppi \over \rho_T + p_T}  \, ,
\label{eqn:xieom}
\end{align}
with $\pi_T$ as the anisotropic stress of the total matter and
$\Delta p_T = \delta p_T^{(T)}$.  Since $V_T = {\cal O}(k_H \zeta)$
we can enforce this condition on large scales by parameterizing a
relationship between $U_e$ and $V_T$ at $k_H \ll 1$
\begin{equation}
\lim_{k_H \ll 1} U_e = -{H^2 \over 12\pi G \rho_e} c_K k_H^2 V_T f_\zeta\,,
\end{equation}
where $f_\zeta(\ln a)$ is a function of time only, {\it i.e.}\ $U_e
= {\cal O}(k_H^3 \zeta)$. Note that $U_e$ is the dark energy
momentum relative to the frame defined by zero matter momentum. The
scaling requirement  is that to first order in $k_H$, the dark
energy and matter rest frames are the same at large scales. Both the
single and multiple scalar field equations exhibit this property
given that $\Delta p_e/\rho_e$ and $\xi$ are ${\cal O}({k_H^2
\zeta})$  (see \cite{Hu04a} Eq.~115 for an explicit expression).
Once $U_e$ and its evolution are determined, $\delta p_e$ follows by
momentum conservation with no singularities encountered as $w_e$
crosses the phantom divide.

The PPF description can be made an exact match at large scales to
any given system of scalar fields with an arbitrary equation of
state evolution  $w_e(\ln a)$ by solving the full equations at $k_H
\rightarrow 0$ and inferring $f_\zeta$ for the evolution of all
other finite $k$ modes.  However for the purpose of obtaining the
correct evolution for the {\it metric} or gravitational potentials,
even this is not necessary as long as
$f_\zeta \simlt {\rho_e /( \rho_T + \rho_e)}.$
By construction, the metric condition $\zeta' = {\cal O}(k_H^2)$ is
satisfied and the specific value chosen just determines the ratio of
the dark energy to matter contributions to the metric fluctuations.
Since ultra large scales where the dark energy is not smooth are
generally probed gravitationally via gravitational redshifts and
perhaps lensing in the future, it suffices for most purposes to
simply take $f_\zeta=0$.

The final piece in the construction is to assure that the dark
energy becomes smooth relative to the matter inside a transition
scale $c_e k_H =1$ while exactly conserving energy and momentum
locally by taking \cite{HuSaw07b, Hu08a}
\begin{equation}
(1 + c_\Gamma^2 k_H^2) [\Gamma' + \Gamma + c_\Gamma^2 k_H^2 \Gamma ] = S\, ,
\label{eqn:gammaeom}
\end{equation}
where
\begin{align}
S&= - {4\pi G \over  H^2} \left[ f_\zeta (\rho_T+p_T) - (\rho_e+p_e)
\right] {V_T \over k_H} \,.
\end{align}
This relation explicitly guarantees that $\Gamma \ll V_T /k_H =
{\cal O}(\Phi)$ for $c_\Gamma k_H \gg 1$.  Comparison with
Eq.~(\ref{eqn:gamma}) shows that this condition requires the dark
energy to be smooth relative to the matter (see
Eq.~\ref{eqn:smooth}).  While the specifics of how rapidly the dark
energy becomes negligible in contributing to gravitational
potentials below this scale depend on the specific form of
Eq.~(\ref{eqn:gammaeom}), the net impact on observable quantities of
this choice is small as we shall see below.

The main task is to calibrate the scale of the transition, {\it
i.e.}\ a relationship between $c_\Gamma$ and $c_e$. We find that
\begin{equation}
 c_\Gamma =   0.4 c_e \,
\end{equation}
matches the evolution of scalar field models.  We show an example
with $w_e=-0.7$ of the evolution of the quantity $(\Phi -\Psi)/2$
that is responsible for gravitational redshifts and lensing in
Fig.~\ref{fig:timeplot}.  Metric evolution for scales $c_e k_H \ll
1$ and $c_e k_H \gg 1$ show exact agreement between the PPF
prescription and the direct scalar field calculation by
construction.   In this model the two limits differ by 44\% in the
fractional change in the gravitational potential during the
acceleration epoch.

In Fig.~\ref{fig:clplot}, we compare the CMB temperature power
spectrum in the PPF approximation to the direct scalar field
calculation for the $w_e=-0.7$ model and a two field model that
approximates $w_e(\ln a) = w_0 + (1-a) w_a$ with $w_0 = -1.15$ and
$w_a=0.5$ \cite{Hu04c}. The latter model has $w_e$ evolves across
the phantom divide.

\section{Discussion}

We have shown that the PPF prescription for describing the evolution
of metric perturbations in an FRW universe is sufficiently general
to encompass multiple scalar field models whose joint equation of
state evolves across the phantom divide at $w_e=-1$. This
description is accurate to well below the cosmic variance limit as
long as the transition scale to relative smoothness is comparable to
the horizon.  Moreover it  is in fact exact for the metric evolution
well above and well below the transition scale.  As such it provides
a well-controlled approximation for any model where the energy and
momentum of the dark energy is separately conserved and features a
transition of this type.

This prescription is useful for the joint analysis of growth and
distance measures of the dark energy, especially those involving
horizon scale perturbations like the integrated Sachs-Wolfe effect
in the CMB.  The CAMB Einstein-Boltzmann package has been altered to
include PPF  \cite{Fangetal08} and a version for the dark energy has
been made publically available
\footnote{\url{http://camb.info/ppf/}}. Potential future uses
include principal component approaches to dark energy constraints
where $w_e$ is allowed to cross the phantom divide multiple times
(e.g.~\cite{HutSta03}).  Here explicit matching to multiple scalar
fields is cumbersome if not  impossible.  The PPF prescription
provides a simple but general approach that explicitly enforces
conservation of energy and momentum and all of the Einstein
equations removing potential ambiguities to the meaning of a
``smooth" dark energy component.

\smallskip

WF is supported by the Initiatives in Science and Engineering
program at Columbia University and by DOE contract
DE-FG02-92ER-40699; WH by DOE contract DE-FG02-90ER-40560, the David
and Lucile Packard Foundation and the KICP under NSF PHY-0114422; AL
by an STFC Advanced Fellowship.

\vfill

\bibliographystyle{arxiv_physrev}

\bibliography{crossing}

\end{document}